# Centre Symmetric Quadruple Pattern: A Novel Descriptor for Facial Image Recognition and Retrieval

[1]Soumendu Chakraborty, [1]Satish Kumar Singh, and [1]Pavan Chakraborty

[1]Indian Institute of Information Technology, Allahabad, India
soum.uit@gmail.com, sk.singh@iiita.ac.in, pavan@iiita.ac.in

**ABSTRACT**

Facial features are defined as the local relationships that exist amongst the pixels of a facial image. Hand-crafted descriptors identify the relationships of the pixels in the local neighbourhood defined by the kernel. Kernel is a two dimensional matrix which is moved across the facial image. Distinctive information captured by the kernel with limited number of pixel achieves satisfactory recognition and retrieval accuracies on facial images taken under constrained environment (controlled variations in light, pose, expressions, and background). To achieve similar accuracies under unconstrained environment local neighbourhood has to be increased, in order to encode more pixels. Increasing local neighbourhood also increases the feature length of the descriptor. In this paper we propose a hand-crafted descriptor namely Centre Symmetric Quadruple Pattern (CSQP), which is structurally symmetric and encodes the facial asymmetry in quadruple space. The proposed descriptor efficiently encodes larger neighbourhood with optimal number of binary bits. It has been shown using average entropy, computed over feature images encoded with the proposed descriptor, that the CSQP captures more meaningful information as compared to state of the art descriptors. The retrieval and recognition accuracies of the proposed descriptor has been compared with state of the art hand-crafted descriptors (CSLBP, CSLTP, LDP, LBP, SLBP and LDGP) on bench mark databases namely; LFW, Colour-FERET, and CASIA-face-v5. Result analysis shows that the proposed descriptor performs well under controlled as well as uncontrolled variations in pose, illumination, background and expressions.

**Keyword:** Local Binary Pattern (LBP), Center Symmetric Local Binary Pattern (CSLBP), Local Derivative Pattern (LDP), Center Symmetric Local Ternary Pattern (CSLTP), Local Directional Gradient Pattern (LDGP), Centre Symmetric Quadruple Pattern (CSQP), face recognition, face retrieval.

## 1. Introduction

Facial image representation using hand-crafted descriptors for recognition and retrieval has been the most researched area in recent past. In general facial image descriptors can be broadly categorized into learning based and hand-crafted descriptors. Local relationships existing amongst pixels of a facial image are used to represent the image uniquely. Such representations are useful where it is required to increase the inter-class dissimilarity and reduce the intra-class dissimilarity. Primary objective of the descriptors such as; Eigen-face [1], Fisher-face [1], variations of PCA [2][3][4][5][6], and Linear Discriminant Analysis (LDA) [7][8][9] is to identify the feature points of the facial images taken under constrained environment. Most recently deep learning framework for facial analysis has been used with large availability of the training data [25-26]. Large training database is required to describe facial features in learning based description. Most of these descriptors are application specific. It is very difficult to generalize these descriptors for various problems such as variations in pose, illumination, expressions and self occlusion [27-28]. Proposed Centre Symmetric Quadruple Pattern (CSQP) is a hand-crafted descriptor encoded for individual images using pixel intensities. Characteristics of CSQP descriptor are different from the characteristics of the VGG model [25-26], which encodes the features of a facial image with Convolutional Neural Networks (CNNs). Local Binary Pattern (LBP) is one of the earliest hand-crafted descriptors used in face recognition [10][30]. A 3 × 3 kernel is used to encode the eight pixels surrounding the centre. LBP performs well under rotation and expression variations. Centre Symmetric Local Binary Pattern (CSLBP) [11], which has a feature length smaller than LBP and its performance is comparable to LBP. Uniform illumination variation significantly degrades the performance of these descriptors. Centre Symmetric Local Ternary Pattern (CSLTP) [12] compensates the uniform illumination by encoding the centre symmetric pixels with +1, 0, and -1. It is a gradient based local descriptor, which works well under controlled illumination variation. Most recently Multi-Block Local Binary Pattern (MB-LBP) has been used to detect pedestrians [29]. Region wise averages are used to define the local descriptor Semi Local Binary Pattern (SLBP) [23], to improve the recognition and retrieval accuracies under scale, noise and illumination variations. There is another class of descriptors defined in the higher order derivative space. Local Directional Gradient Pattern (LDGP) [15] is one of the most recent higher order descriptor, which shows improvements over Local Derivative Pattern (LDP) [13], and Local Vector Pattern (LVP) [14] with respect to time and achieves comparable recognition rates. CSQP encodes the pixels in the larger neighbourhood, as under unconstrained variations in illumination, background and pose the increasing neighbourhood reduces the intra-class dissimilarity [18].

The organization of the rest of the paper is as follows. Section 2 elaborates the motivation and proposition of the descriptor. Various experiments have been performed and the obtained results are compared with the state of the art descriptors in section 3.

The work reported through this paper is concluded in section 4.

## 2. Proposed Centre Symmetric Quadruple Pattern

*2.1 Motivation*

The facial images are by and large symmetric across vertical axis and asymmetric across horizontal axis. Most of the hand-crafted descriptors such as LBP, SLBP encodes the asymmetry and symmetry of a facial image with respect to the centre pixel of the kernel. In a facial image the pixels in the upper as well as the lower half of the image more often than not show very little variations, which could be useful for facial image representation. Hence some of the descriptors such as CSLBP and CSLTP encode diagonally opposite pixels across the centre to capture the meaning full asymmetry. The local neighborhood of the facial image shows significant diagonal asymmetry at certain transition points. One of these transition points is shown in Fig.1. The proposed descriptor has been designed to capture such dissimilarities in a facial image. Under unconstrained variations in illumination, background, and pose encoded asymmetry tend to increase the intra-class distance due to the small size of the local neighborhood. To overcome these problems CSQP has been designed in such a way that it captures meaningful asymmetry in the diagonally opposite quadruple space with larger neighborhood. To verify that the proposed descriptor captures meaningful information, the average entropy of the proposed and the state of the art descriptors have been computed over CASIA-face-v5 [19] database. The average entropy of CSLBP, CSLTP, LDP, LBP, SLBP, LDGP and CSQP are 3.44, 1.92, 4.63, 4.06, 2.85, 3.27 and 6.94. This entropy values clearly show that the proposed method contains more information as compared to the state of the art. Result analysis confer that these additional distinctive relationships are useful enough to significantly increase the accuracy of the proposed descriptor under constrained and unconstrained environment.

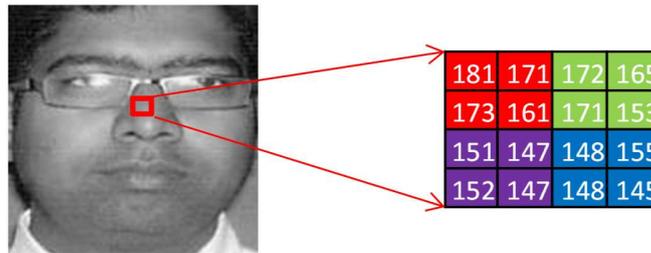

Fig. 1. Local neighborhood around nose showing horizontal symmetry and diagonal asymmetry.

*2.2 Major contribution*

The feature representation under constrained variations in pose, illumination and background requires only a small number pixel to be encoded in the vicinity of the reference pixel. CSLBP, CSLTP and LBP are some of the examples of such descriptors which encode only a limited number of pixels as shown in Table 1. Facial images taken in the real world environment are more difficult to represent with limited local neighborhood pixels for accurate recognition and retrieval. There exist descriptors such as LDP, LGHP [18] which encode pixels in the larger neighborhood. However, feature lengths of these descriptors are the major cause of concern. Large feature lengths adversely affect the feature extraction time and match time of the recognition system. To improve the recognition and retrieval rates, we propose a novel scheme CSQP to encode pixels in the larger neighborhood of $4 \times 4$ block. The proposed method encodes 16 pixels with only an 8 bit pattern. The proposed descriptor effectively captures diagonal asymmetry and vertical symmetry in a facial image. CSQP template computes distinctive information from a facial image as shown by the entropy values given in the previous section. CSQP with larger neighborhood reduces the intra-class distances of the facial images with severe variations in environmental conditions. Even the computation complexity of the descriptor is in the order of the size of the image, which is comparable to the most efficient class of image descriptors.

*2.3 Proposed descriptor*

CSQP divides the local kernel of size $4 \times 4$ into 4 sub-blocks of size $2 \times 2$ as shown in Fig.2. Pixels of the diagonally opposite sub-blocks are compared to generate a binary pattern of 8 bits.

Fig. 2. Template showing the local neighborhood of the reference pixel $I_{i,j}$.

Pixels of the red sub-block are compared with the pixels of the blue sub-block as shown in (1) using encoding function given in (3) and their weighted sub is computed. Similarly, pixels of the green sub-block are compared with the pixels of the purple sub-block and the weighted sum of the resulting binary code is computed as shown in (2).

$$A_{i,j}^1 = 2^7 \times C(I_{i,j}, I_{i+2,j+2}) + 2^6 \times C(I_{i,j+1}, I_{i+2,j+3}) + 2^5 \times C(I_{i+1,j}, I_{i+3,j+2}) + 2^4 \times C(I_{i+1,j+1}, I_{i+3,j+3}) \quad (1)$$

$$A_{i,j}^2 = 2^3 \times C(I_{i,j+2}, I_{i+2,j}) + 2^2 \times C(I_{i,j+3}, I_{i+2,j+1}) + 2^1 \times C(I_{i+1,j+2}, I_{i+3,j}) + 2^0 \times C(I_{i+1,j+3}, I_{i+3,j+1}) \quad (2)$$

Where $i = 1,2..M - 3, j = 1,2 ... N - 3$ and $C$ is the encoding function defined as

$$C(E,F) = \begin{cases} 0, & if E \leq F \\ 1, & else \end{cases} \quad (3)$$

$A_{i,j}^1$ and $A_{i,j}^2$ are combined to compute the decimal equivalent of the 8 bit pattern generated from all blocks in (4).

$$A = A_{i,j}^1 + A_{i,j}^2 \quad (4)$$

Finally the histogram of the feature image computed using (4) is computed in (5), which is feature computed by the CSQP.

$$CSQP = \{H_A\} \quad (5)$$

$\chi^2$ distance [16] is used to measure the similarity between two histograms. Similarity measure $S_{\chi^2}(.,.)$ is defined as

$$S_{\chi^2}(X,Y) = \frac{1}{2}\sum_{i=0}^{q} \frac{(x_i - y_i)^2}{(x_i + y_i)} \quad (6)$$

where $S_{\chi^2}(X,Y)$ is the $\chi^2$ distance computed on two vectors $X = (x_1, ..., x_q)$ and $Y = (y_1, ..., y_q)$. Nearest one neighbor (1 NN) classifier is used as reported in [13] to compute the minimum $\chi^2$ distance between the probe and the gallery images. As similar regions of the probe and gallery images are effectively identified by 1NN classifier with optimal computational cost [13].

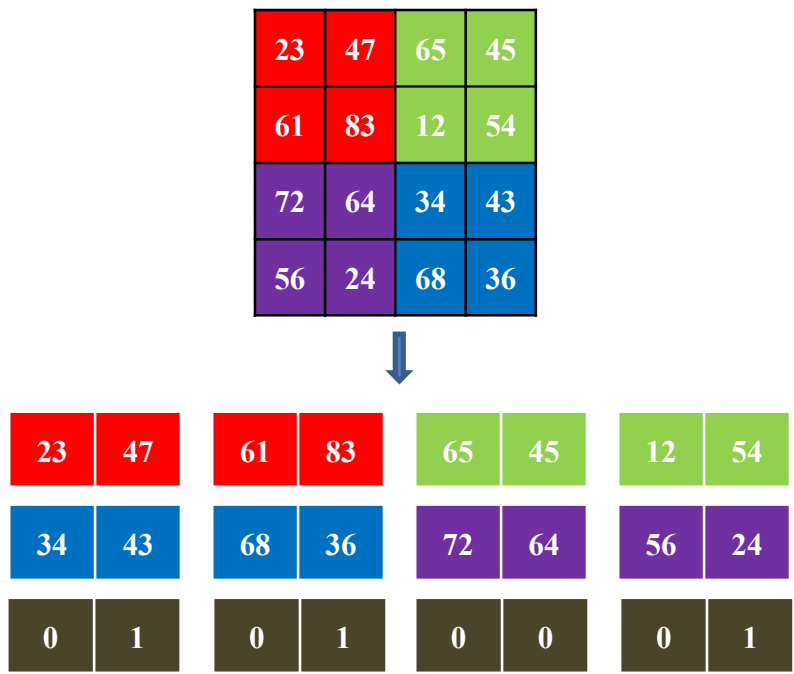

Fig. 3. Example showing the encoding scheme of CSQP.

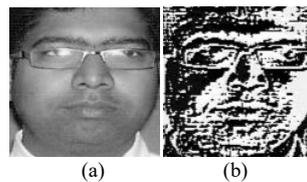

(a)　　　　　(b)
Fig. 4. (a) Original image, (b) CSQP Feature image

A sample image and the encoding structure have been depicted in Fig, 3. Encoding function (3) has been used to generate the 4 bit pattern by comparing the pixel intensities in red sub-block with the pixel intensities in blue sub-block. The resulting 4 bit pattern has been shown in gray at left side of the bottom of the Fig. 3. Similarly, another 4 bit pattern has been computed by comparing the pixel intensities in green sub-block with the pixel intensities in purple sub-block. The resulting 4 bit pattern has been shown at right side of the bottom of the Fig. 3. Finally these two binary patterns are converted to equivalent decimal values and combined together to generate the feature image.

These feature image is shown in Fig.4(b) for the original image shown in Fig.4(a). Feature image clearly shows that the descriptor captures distinctive complementary relationships that exist in the local neighborhood of the reference pixel.

Major differences and novelty of the proposed descriptor are summerised as follows

1. SLBP computes the averages of the 2×2 overlapping blocks and then computes LBP over such averages. Significant loss of information caused by the average taken over 2×2 overlapping blocks degrades the performance of SLBP. As opposed to SLBP, proposed CSQP directly compares 2×2 overlapping blocks in diagonally opposite directions to capture the asymmetry at several transition points. The feature lengths of SLBP and CSQP are same. However, CSQP shows significant gains in retrieval and recognition accuracy across all databases as shown in section 3.

2. LDGP is a descriptor, which is defined in the higher order derivative space, whereas CSQP encodes pixels of the original image. LDGP compares only the reference pixel across different derivative space, whereas CSQP compares block of pixels in the diagonally opposite directions. Even though the length of LDGP is less than the length of CSQP, the improved accuracy of about 8%-10% of CSQP makes the contribution of this descriptor significant.

3. LDP is another derivative based descriptor which compares the 3×3 neighborhood in four different derivative spaces. Hence it is evident that LDP is structurally more close to LDGP, LBP and CSLBP, as the local neighborhood of these descriptors are similar. The encoding structure, as well as the local neighborhood of CSQP is different from LDP. CSQP captures block wise dissimilarity in diagonally opposite regions. On most challenging databases namely; LFW and FERET, the retrieval and recognition accuracy of LDP are close (still inferior) to the accuracy of CSQP. The feature length of LDP (32 bits) is four times the length of CSQP (8 bits). This reduction in feature length is significant. If we compare the length and the accuracy of LDP with CSQP the superiority of CSQP is noticeably quite significant.

4. The descriptor which is structurally close enough to proposed CSQP is CSLBP. CSLBP encodes the local neighborhood similar to the local neighborhood proposed in LBP. CSLBP only compares pair of pixels in horizontally, vertically, and diagonally opposite directions whereas CSQP compares entire blocks in the diagonally opposite direction. The performance of CSLBP is comparable (still inferior) to CSQP only on one database, whereas CSQP completely outperforms CSLBP on more challenging databases namely; FERET and LFW.

The method of description of the facial image using block wise encoding of a neighborhood, which is different and effective as compared to the neighborhood used in LBP, LDP, LDGP, SLBP, CSLBP itself is novel.

*2.4 Feature image analysis*

The effect of variations in illumination, pose with self-occlusion on the feature images has been analyzed in this section. Fig.5(a-d) shows the original facial images with unconstrained variations in pose and illumination. Corresponding feature images are shown in Fig.5(e-h). It is visually evident that even with these severe variations the posed descriptor has been able to extract the dominant visible features such as eyes, nose, lips etc.

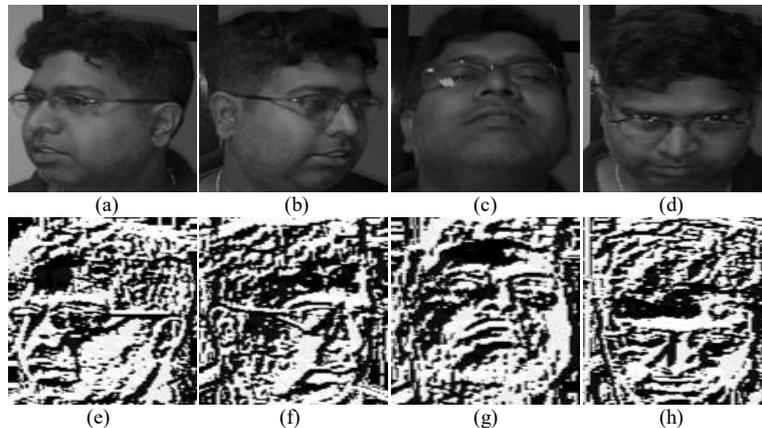

Fig. 5. Visual representation of CSQP feature images for varying pose and illumination: (a) original image left profile, (b) original image right profile, (c) original image up profile, (d) original image down profile (e) Feature image left profile, (f) Feature image right profile, (g) Feature image up profile, (h) Feature image down profile.

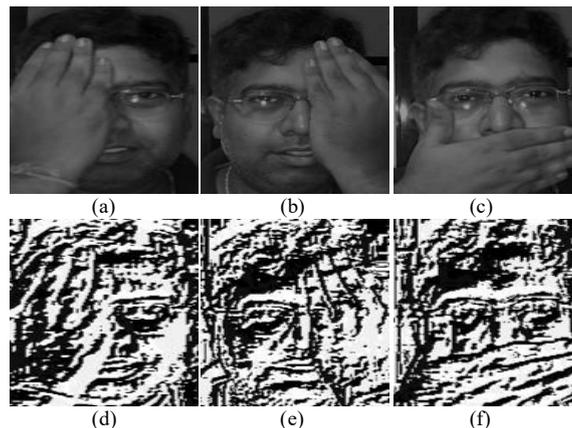

Fig. 6. Visual representation of CSQP feature images with self occlusion: (a) original image with occluded right eye and some portion of the lips, (b) original image with occluded left eye and some portion of the lips, (c) original image with occluded lips, (d) Feature image with occluded right eye and some portion of the lips, (e) Feature image with occluded left eye and some portion of the lips, (f) Feature image with occluded lips.

Several experiments were conducted on a number of facial images with self-occlusion and illumination, expression variations. Some of the sample images are shown in Fig.6(a-c). Feature images of these samples are shown in Fig.6(d-f). Visible features under self-occlusion are computed correctly by the proposed descriptor. Hence it can be concluded that the proposed descriptor should be able to achieve comparable performance under self-occlusion.

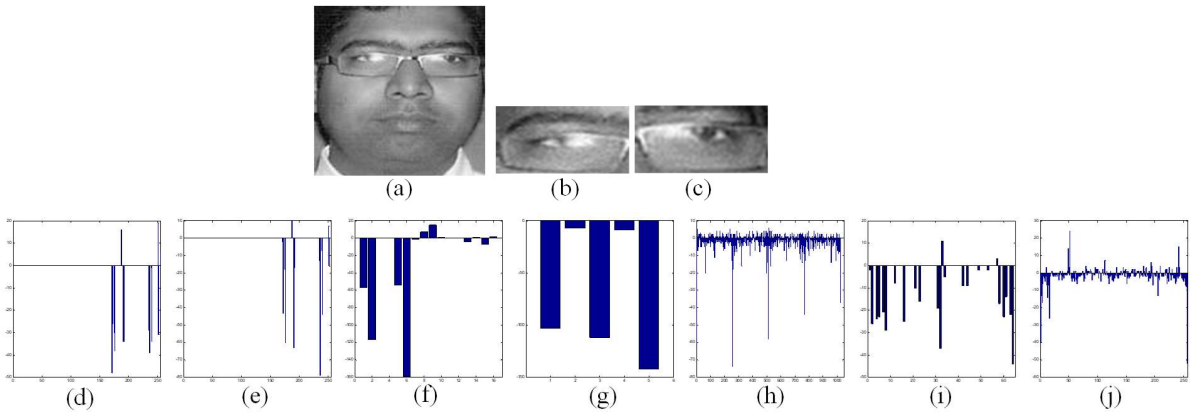

Fig. 7. (a) Original image, (b) Right eye, (c) Left eye. Difference histogram of left and right eye computed from (d) LBP ($\sigma^2 = 67.71$), (e) SLBP ($\sigma^2 = 75.39$) (f) CSLBP ($\sigma^2 = 2452$), (g) CSLTP ($\sigma^2 = 3498$), (h) LDP ($\sigma^2 = 48.19$), (i) LDGP ($\sigma^2 = 116.57$), and (j) CSQP ($\sigma^2 = 31.29$).

To show that the proposed descriptor effectively encodes the vertical symmetry of a facial image features of the left and right eye of a sample image are computed and their difference histograms are shown in Fig.7. Uniform distribution of the positive and negative errors computed by the CSQP as shown in Fig.7(j) illustrates that the proposed descriptor captures the similarity of the left and right half of a facial image. The variance ($\sigma^2$) of the difference histograms obtained using LBP, CSLBP, CSLTP, SLBP, LDP, LDGP, and CSQP are 67.71, 2452, 3498, 75.39, 48.19, 116.57, and 31.29. Minimal cumulative error energy of the difference histogram enables the proposed descriptor to reduce the intra-class distance of the probe image.

**3 performance analysis**

The performance of the proposed descriptor CSQP has been analyzed in recognition as well as retrieval framework. Different performance measures, such as Average Retrieval Precision (ARP), Average Recall Rate (ARR), F-Score (F) and Average Normalized Modified Retrieval Rank (ANMRR) [18][24], are used to evaluate and compare the proposed descriptor. Precision is defined as the number of relevant images retrieved out of the total number of retrieved images and Recall is defined as the total number of relevant images retrieved [18][24]. ARP and ARR are computed by taking the means of the precision and recall respectively. F-Score is computed as the Harmonic mean of the corresponding ARP and ARR. The F-Score depicted in all the figures is the average of the F-Score computed per class.

ANMRR defined in [17][24] is used to measure the performance of the descriptors based on the rank of the retrieved images. Low ANMRR indicates that the images retrieved by the descriptor are highly relevant to the queried image and higher value of ANMRR indicates that most of the top ranked retrieved images by the descriptors are not relevant to the queried image. ANMRR is computed for the maximum number of images in a particular category of the database.

Performance of the proposed method has been analyzed on the latest and most challenging facial image databases namely: CASIA-Face-V5-Cropped [19], Color FERET [20] [21], and LFW [22]. Lengths of the proposed and other state of the art descriptors are shown in Table 1. Table 1 also shows the number of neighborhood pixels encoded by the descriptor. Length of the proposed descriptor is half of the length of LDP and it encodes twice the number of pixels encoded by LDP. Even though the lengths of the descriptors namely CSLBP, CSLTP, LBP, and LDGP are less than the length of CSQP, the number of neighborhood pixels encoded by these descriptors is less than the half of the pixels encoded by CSQP.

*3.1 Performance analysis on CASIA-Face-V5-Cropped database*

"Portions of the research in this paper use the CASIA-FaceV5 collected by the Chinese Academy of Sciences' Institute of Automation (CASIA)" [19]. CASIA-Face-5.0 database contains 5 color images each of the 500 individuals. Images are captured with intra-class variations such as illumination, pose, expressions, eye-glasses, and imaging distance [19].

The comparative ARP and ARR values shown in Fig.7(a-b), illustrates that the retrieval rates of CSQP are comparable to the retrieval rates of CSLBP for rank 2. CSQP performs even better increasing number of retrieved images. CSQP achieves approximately 0.2% improvement over state of the art image descriptors. Better average F-Score shown in Fig.7(c) demonstrates consistent improvement of 1% over CSLBP. CSQP outperforms most recent descriptors such as LDGP, LDP, and SLBP. It shows 2%, 4%, and 3% improvement over LDGP, LDP, and SLBP respectively. Lowest ANMRR of CSQP shown in Fig.7(d) verify that the proposed descriptors retrieves more images with low rank (more relevant). Hence it can be concluded from over all analysis of the descriptor on CASIA-Face database that it outperforms state of the art descriptors with respect to retrieval rates.

## 3.2 Performance analysis on Color FERET database

"Portions of the research in this paper use the FERET database of facial images collected under the FERET program, sponsored by the DOD Counterdrug Technology Development Program Office". Color-FERET database is one of the most challenging facial image databases with severe variations in pose and expression. The color FERET database contains 11,338 facial images of 994 individuals at different orientations. There are 13 different poses used in the images of the database [20][21].

Table 1: Length of the descriptors

| Descriptor (year) | Length (bits) | Length (bins) | Pixels Encoded |
|---|---|---|---|
| CSLBP (2009) | 4 | 16 | 8 |
| CSLTP (2010) | - | 9 | 8 |
| LDGP (2017) | 6 | 64 | 4 |
| LBP (1996) | 8 | 256 | 8 |
| SLBP (2015) | 8 | 256 | 4×4 |
| LDP (2010) | 8×4 | 256×4 | 8 |
| CSQP (Proposed) | 8 | 256 | 4×4 |

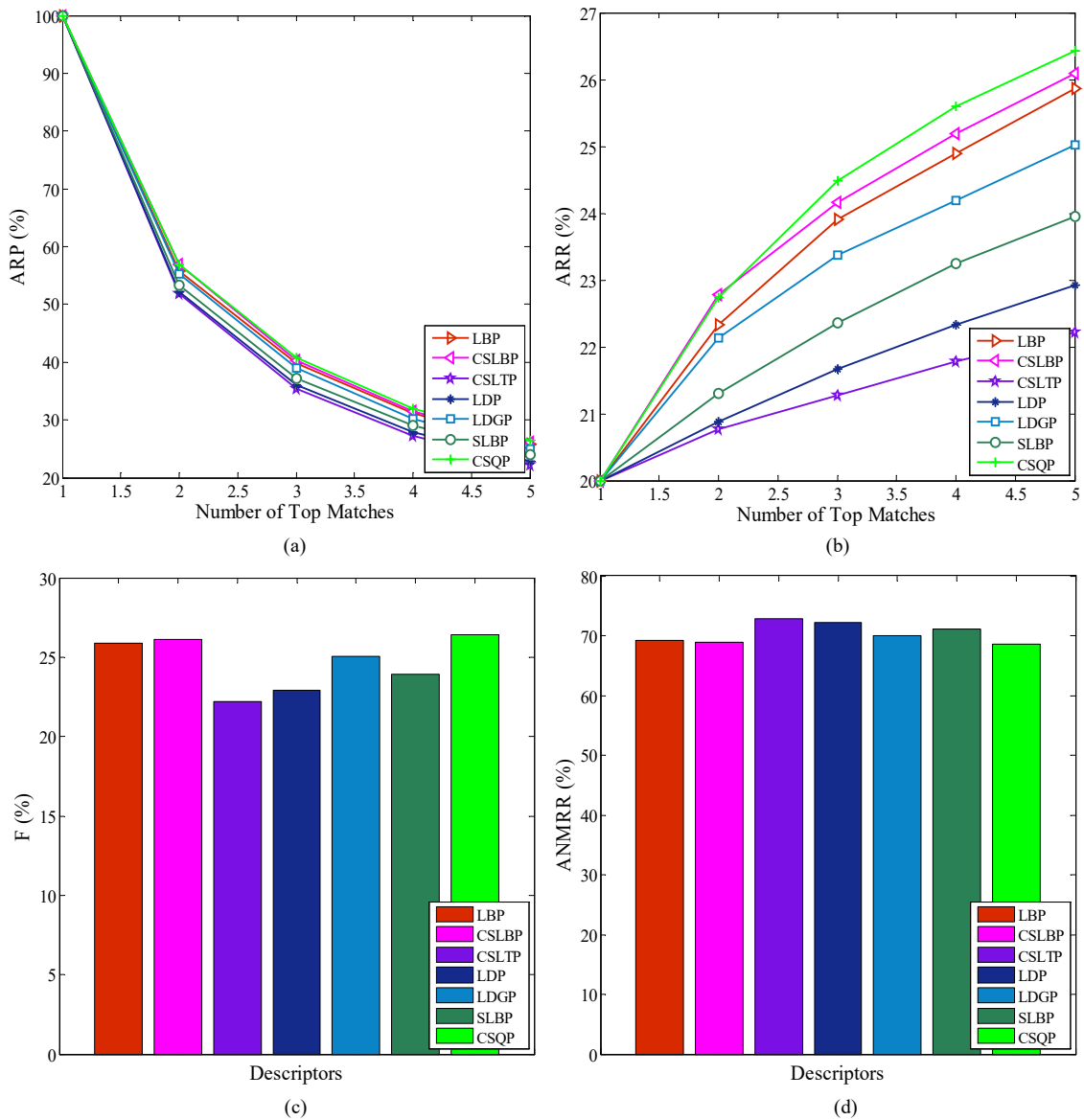

(a)  (b)  (c)  (d)

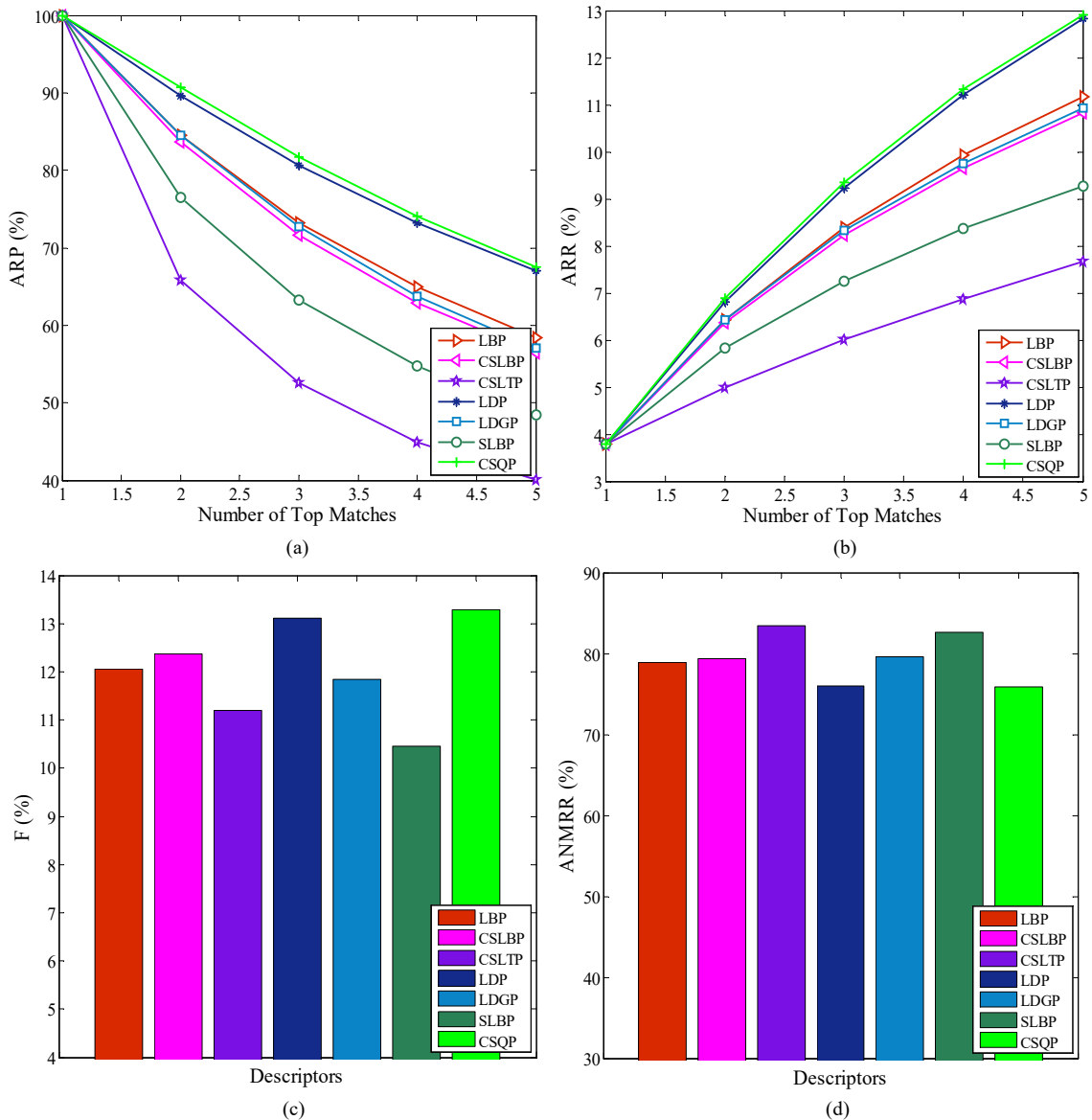

Fig. 8. (a) ARP, (b) ARR, (c) F-Score, and (d) ANMRR. computed over CASIA-Face database.

Fig. 9. (a) ARP, (b) ARR, (c) F-Score, and (d) ANMRR. computed over Color-FERET database.

Feature description under unconstrained environment is more challenging as compared to the feature description under constrained variations. Color-FERET database is an ideal database to test the robustness of any hand-crafted descriptor under unconstrained environment. Retrieval rates (ARP and ARR) of CSQP are consistently higher than the state of the art descriptors such as LDGP, LDP, SLBP with increasing number of retrieved images as shown in Fig.8(a-b). CSQP achieves 1%, 10% and 20% better ARP over LDP, LDGP, and SLBP respectively. Retrieval rates of LDP and CSQP are showing comparable performance. However, smaller length of CSQP makes it a clear winner over LDP. Higher average F-Score of CSQP shown in Fig.8(c), illustrates consistent and better performance over LDP, LDGP, and SLBP. Better ANMRR of CSQP shown in Fig.8(d) validates that the images retrieved using CSQP are the closest to queried image from the same class (images with low rank are closer to queried image).

*3.3 Performance analysis on LFW database*

There are 13,233 color facial images of 5,749 individuals in LFW database [22]. 1680 individuals have two or more images and rest of the individuals have only one image. As the proposed descriptor has been designed to work under unconstrained environment, we test the proposed descriptors over this database.

The descriptor is tested on LFW database as it is one of the largest and most challenging databases of facial images taken under uncontrolled real world environment. Performance of the proposed descriptor has been evaluated on LFW to show its robustness against unconstrained variations in pose, illumination, and expression.

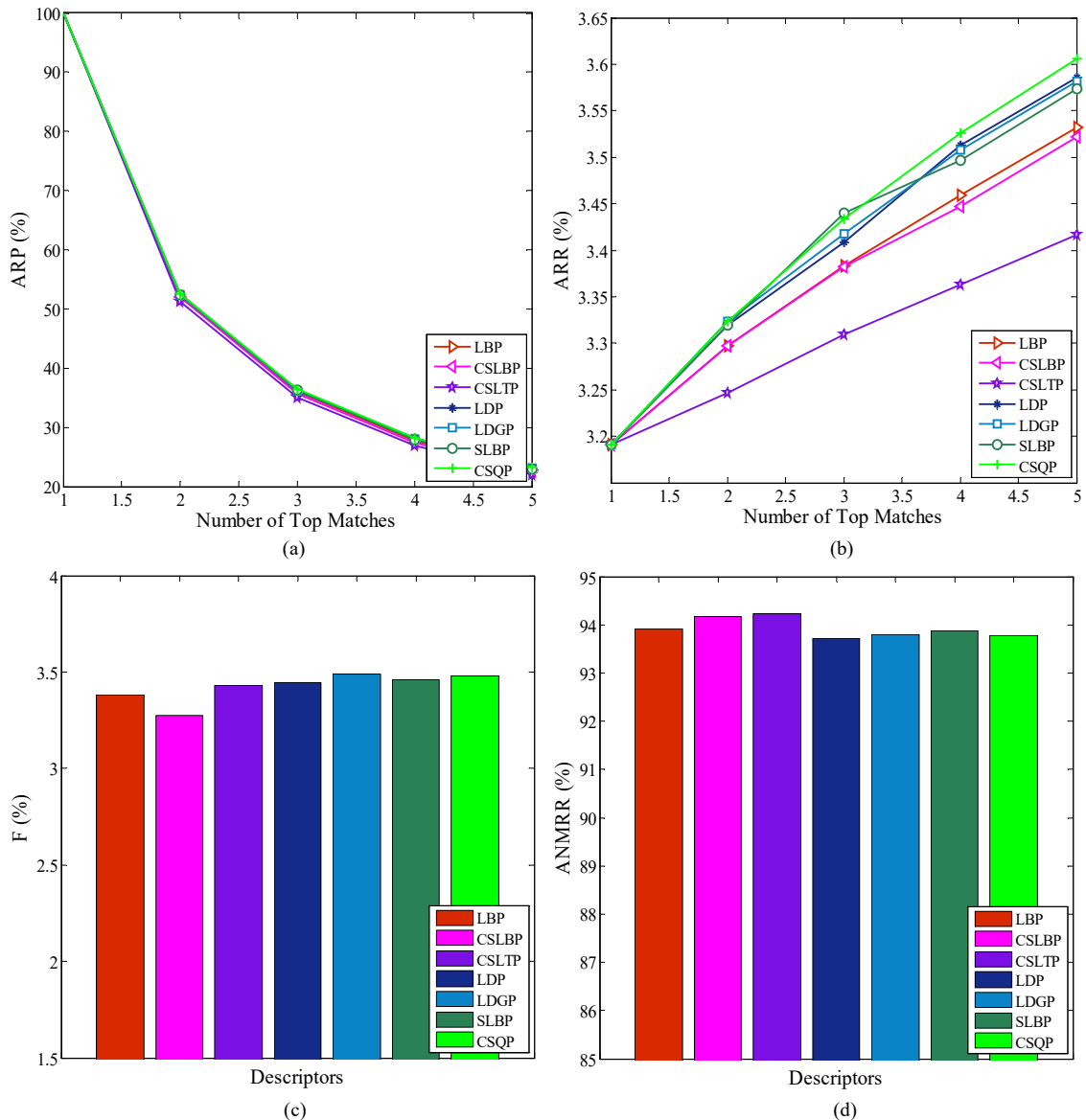

Fig. 10. (a) ARP, (b) ARR, (c) F-Score, and (d) ANMRR. computed over LFW database.

Retrieval accuracy of CSQP is comparable to SLBP on LFW database as shown in Fig.9(a-b). The performance of CSQP improves with increasing retreived images. The average F-Score of CSQP shown in Fig.9(c) is better than SLBP and comparable to LDGP. Hence it can be concluded that the proposed descriptor retrieves more relevant images than SLBP. As illustrated by comparably lowest ANMRR of CSQP shown in Fig.9(d), most of the images retrieved by CSQP are the low rank images (low rank images are closer to queried image).

*3.4 Comparative analysis between CSQP and LQPAT*

Local Quaduple Pattern (LQPAT) [17] encodes the pixels of four 2×2 blocks in clockwise direction, where as CSQP encodes diagonally opposite 2×2 blocks to achieve reduced feature length. Feature length of CSQP is half the length of LQPAT. The recognition and retrieval accuracies of CSQP and LQPAT are almost same as shown in Table 2. It is desirable that the maximum ARP and ARR should be at rank 1 for disjoined sets of probe and gallery. CSQP achieves maximum ARP and ARR at rank 1 which is almost equal to the ARP and ARR of LQPAT.The difference in retreival accuracy is as low as 0.06% and 0.03% on CASIA and LFW databases.

*3.5 Performance analysis in recognition framework*

Recognition rates are computed by taking each image in the database as probe and rest of the images as gallery. If there are $N$ images in the database then each image is taken as probe in turn and rest $(N-1)$ images are taken as gallery. The distance between probe feature and gallery feature is computed using $\chi^2$ distance. There are $(N-1)$ distances for each probe. The gallery image with lowest distance is given the lowest rank. If the gallery image with lowest rank belongs to the same class as the class of the probe image then it is taken as a match. If $N_m$ be the total number of matches, then the recognition rate ($R_R$) of a descriptor is computed as

$$R_R = (N_m/N) * 100 \qquad (7)$$

The Cumulative Match Characteristics (CMCs) for different data bases are shown in Fig.10. For CMC a match is taken, if the class of the probe image matches with the class of at least one gallery image with rank less than or equal to the maximum rank specified.

Table 2: Comparative analysis of CSQP and LQPAT

| Database | Length (bits) | | Length (bins) | | Maximum ARP (%) | | Maximum ARR (%) | | F-Score (%) | | ANMRR (%) | |
|---|---|---|---|---|---|---|---|---|---|---|---|---|
| | CSQP | LQPAT | CSQP | LQPAT | CSQP | LQPAT | CSQP | LQPAT | CSQP | LQPAT | CSQP | LQPAT |
| CASIA-FACE-V5 | 8 | 16 | 256 | 512 | 58.03 | 58.09 | 26.4 | 27 | 25.3 | 25.4 | 69 | 68.8 |
| Color-FERET | | | | | 91 | 92 | 13 | 13.2 | 13.4 | 13.5 | 76 | 75 |
| LFW | | | | | 53 | 53.02 | 3.6 | 3.63 | 3.45 | 3.5 | 94 | 93.6 |

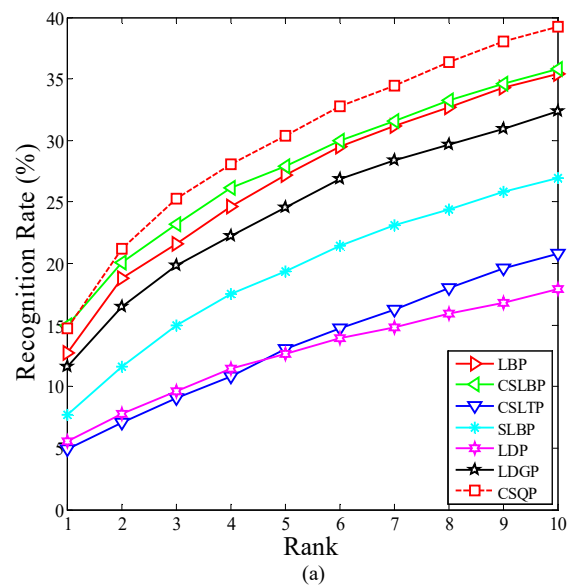

(a)

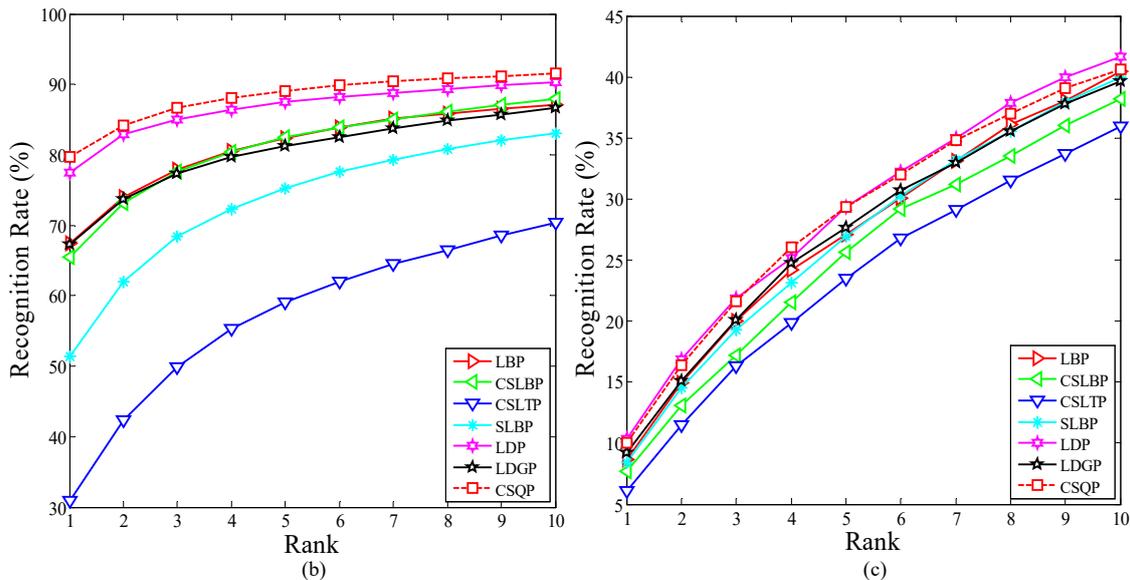

Fig. 11. CMC for different descriptors and CSQP on databases (a) CASIA-Face-V5-Cropped, (b) Color-FERET, and (c) LFW.

The proposed method shows consistent improvement in recognition rate with increasing ranks. CSQP shows 3% and 2% improvement over its nearest counterpart on Color-FERET and CASIA-Face-V5-Cropped databases respectively. The proposed method shows recognition rates comparable LDP up to 7 retrieved images. As the length of CSQP is half the length of LDP, it can be said that CSQP outperforms LDP. Significant and steady improvement shown by the proposed method on the most challenging databases illustrates the robustness of the proposed descriptor against pose, illumination, background and expression variations.

The performance analysis conducted in recognition framework shows that the proposed descriptor achieves consistent improvement over other state of the art descriptors with varying size of gallery. It also shows that the proposed descriptor attains better recognition rates with increasing size of gallery even on most challenging databases.

*3.6 Complexity Analysis*

The computational complexity of the proposed descriptor is comparable with the complexity of the state of the art descriptors SLBP, LDP, LDGP etc. In the local neighborhood the proposed descriptor requires four comparisons to compute a four bit pattern. There are two such patterns which require 8 comparisons. Hence the proposed descriptor takes 8 comparisons to compute the micropattern in the local neighborhood of the reference pixel. The decimal conversions as shown in (1)-(2) require 6 additions and 8 multiplications which add up to 14 fundamental operations. Hence the total number of operations required to compute the feature image from an image of size $M \times N$ is $22 \times M \times N$. Therefore the computational complexity of the descriptor is $O(M \times N)$ which is computationally comparable with the complexity of the state of the art descriptors.

**4 Conclusions**

Hand-crafted descriptors are designed to recognize facial images under pose, illumination, background variations. Most of these descriptors are proposed for controlled variations. To cope with the severe variations in the real world environmental conditions the local neighborhood of these descriptors need to accommodate more pixels, which increases the feature length of the descriptor. Proposed CSQP overcomes this drawback of the existing descriptors by encoding the pixels in the local neighborhood in quadruple space. CSQP computes an eight bit pattern from 16 pixels in the local neighborhood. This effective encoding scheme increases the retrieval and recognition accuracy under constrained and unconstrained pose, illumination, background variations.